\documentclass[conference]{IEEEtran}
\usepackage[utf8]{inputenc}
\usepackage[T1]{fontenc}

\usepackage{fancyhdr}
\usepackage[T1]{fontenc}
\usepackage{graphicx}
\usepackage{cite}
\usepackage{tikz}
\usepackage{amsmath}

\usepackage[toc,acronym]{glossaries} 
\usepackage{lipsum}
\usepackage{xspace}

\usepackage[linesnumbered,ruled,vlined]{algorithm2e}
\usepackage{amssymb}
\usepackage{dirtytalk}
\usepackage{subcaption}
\usepackage{algpseudocode}
\usepackage{indentfirst}
\usepackage[abbreviations]{foreign} 

\usepackage{enumitem} 
\usepackage{caption}
\usepackage{multirow}
\usepackage{pdfpages}

\usepackage{array}

\usepackage[draft,bookmarks=false]{hyperref}
\usepackage{amssymb}
\usepackage{booktabs}
\usepackage{nomencl}
\makenomenclature
\usepackage{etoolbox}

\renewcommand\nomgroup[1]{%
    \item[\bfseries
                \ifstrequal{#1}{N}{Nodes/Participants in a distributed system}{%
                    \ifstrequal{#1}{S}{Sketching parameters in large scale system}{}}%
          ]}

\newcommand{\mysubfigwidth}{0.255\textwidth}

\renewcommand{\arraystretch}{1}     

\newcolumntype{P}[1]{>{\centering\arraybackslash}m{#1}}

\newcommand\blfootnote[1]{%
  \begingroup
  \renewcommand\thefootnote{}%
  \footnote{#1}%
  \addtocounter{footnote}{-1}%
  \endgroup
}

\author{
\IEEEauthorblockN{Augusta Mukam , Joachim Bruneau-Queyreix, Laurent Reveillere}
\IEEEauthorblockA{Univ. Bordeaux, CNRS, Bordeaux INP, LaBRI, UMR 5800, F-33400 Talence, France}
}
\begin{document}

\bstctlcite{BSTcontrol}

\title{Hopper: Bounded-Memory Collaborative Debiasing for Byzantine-Tolerant Peer Sampling}

\maketitle
\blfootnote{The research leading to these results has received funding from the French National Research Agency (ANR) under grant ANR-21-CE25-0021-03}

\begin{abstract}
    Byzantine-tolerant peer sampling relies on continuously refreshed views, yet an adversary can bias the identifier streams used to construct them. Frequency-aware debiasing downweights overrepresented identifiers, but existing designs rely on cumulative per-identifier counts. We show that even exact, unbounded counters fail under a delayed balanced attack, in which a long benign prefix masks a subsequent adversarial frequency shift.
    We introduce Hopper, a bounded-memory debiasing protocol for Byzantine-tolerant peer sampling. We identify the stream-estimation properties required for debiasing and select BitMatcher as the estimator that best preserves adversarial frequency structure among the evaluated alternatives. Hopper adds BMDecay, a saturation-triggered decay and reconstruction mechanism that keeps this signal fresh over long executions. Hopper also supports trusted collaboration through authenticated fingerprint-aware reconstruction and role-specific debiasing. Experiments show that Hopper recovers from delayed attacks faster than when relying on BitMatcher, and debiaising as well as non-debiasing baselines under a fixed memory budget. Trusted collaboration reduces post-attack pollution peaks but creates a re-identification trade-off at high trusted-node densities. These results show the importance of occurence freshness, rather than exact counting alone, as a key requirement for practical frequency-aware Byzantine peer sampling.
\end{abstract}

\section{Introduction}
\label{sec:introduction}

Large decentralized systems cannot maintain a complete membership list at every node. A \emph{peer-sampling protocol} instead gives each node a small, continuously refreshed set of identifiers, called its \emph{view}, from which it selects communication partners~\cite{peersampling:2004,jelasityGossipbased2007,voulgarisCYCLON2005,tolgyesi2009adaptive}. The objective is for every identifier to appear with approximately the same probability.
In adversarial conditions, with a fraction $f$ of Byzantine nodes, the adversary coordinates exchanges to overrepresent Byzantine identifiers in the views of correct nodes, making correct nodes contact the adversary more often and potentially enabling eclipse attacks~\cite{Heilman:2015,eclipse:2006}.
Resilience to such Byzantine manipulation is therefore central to the security of applications that rely on peer sampling, \eg, consensus protocols~\cite{korkmaz2022alder, yahyaoui2024tolerating,gilad2017algorand,amores2026analysis}, decentralized learning~\cite{blade:2025, touat2025exposing}, information dissemination~\cite{korkmaz2022idagossip,infodiss, yahyaoui2025hermes},
and service discovery~\cite{discng:2024,constellation:2026,lift:2025}.


Byzantine-tolerant peer-sampling protocols seek near-uniform view composition despite such manipulation~\cite{Bortnikov:2008,basalt:2023,Pigaglio:2022,aupe:2024}.
BRAHMS~\cite{Bortnikov:2008} is a foundational peer-sampling protocol resilient to Byzantine attacks.
In each round, a node \emph{pushes} its identifier and \emph{pulls} selected peers' views, then rebuilds its view from three subviews: a push subview sampled from received pushes, a pull subview sampled from received pull replies, and a persistent min-wise history subview~\cite{Bortnikov:2008}. Under BRAHMS's assumptions, the history subview converges toward a Byzantine fraction bounded by $f$ once the node has received enough distinct identifiers.
A canonical adversarial strategy against BRAHMS is a \emph{balanced attack}, which spreads Byzantine identifiers evenly through push messages while Byzantine nodes return pull replies containing only Byzantine identifiers, thereby biasing both \emph{push} and \emph{pull} input streams.
AUPE~\cite{aupe:2024} addresses this weakness by placing a \emph{Set Cleanser} before BRAHMS constructs its push and pull subviews.
Inspired by work on adaptive sampling~\cite{anceaume2013uniform}, AUPE maintains a cumulative occurrence counter for each received identifier and inserts an identifier into a sample memory with probability inversely proportional to its exact occurrence count before using it to construct the push and pull subviews.
This approach debiases the input streams by reducing the sample memory insertion probability of identifiers that have been overrepresented in the past, thereby reducing the view pollution caused by Byzantine nodes.
Additionally, AUPE allows trusted nodes protected by the hardware and code-integrity guarantees of trusted execution environments (TEEs) to combine occurrence-count observations and improve debiasing.

AUPE fundamentally leaves two assumptions unresolved for long-lived systems.
First, exact tracking requires one entry per observed identifier.
Consequently, memory usage grows with an unknown population that can greatly exceed the view size.
Each exact occurrence count also grows throughout the execution, requiring progressively wider counters or explicit handling of fixed-width counter saturation.
AUPE suggests relying on a bounded-memory Count-Min Sketch (CMS) occurrence estimator, but the collisions that bound its memory footprint systematically overestimate occurrences and may compress the occurrence count differences that drive the identifier insertion rule.

Second, cumulative counters weigh old and recent observations equally.
This enables a \emph{delayed balanced attack}, in which Byzantine nodes behave normally for a long period, accumulate large and similar occurrence counts, and only then launch a balanced attack.
Byzantine identifiers subsequently dominate the streams received by each node, but their cumulative counts and insertion probabilities remain close to those of correct identifiers.
Exact counting is therefore unable to prevent pollution of correct nodes' views because the failure comes from stale history, not estimation error.

Figure~\ref{fig:delayed-baselines} experimentally validates this limitation.
We emulate a 1,000-node system with 20-identifier views.
Byzantine nodes behave correctly for the first 10K rounds and then launch a balanced attack.
With a Byzantine fraction of $f=20\%$, AUPE's view pollution rises from the uniform target of 0.20 to approximately 0.68 at attack onset and remains above the target fraction $f$ throughout the post-attack interval.
Across Byzantine fractions, AUPE improves over alternative protocols such as BRAHMS and BASALT~\cite{basalt:2023} but does not restore uniform sampling within the following 10K rounds, a considerable delay for a long-lived system in which rounds are typically measured in minutes.
This result shows that effective long-term debiasing in peer-sampling protocols requires a bounded-memory solution that remains responsive to changes in the recent identifier-occurrence distribution.

\renewcommand{\mysubfigwidth}{0.16\textwidth}
\begin{figure}[!t]
  \centering
  \begin{subfigure}{\mysubfigwidth}
    \includegraphics[width=\linewidth]{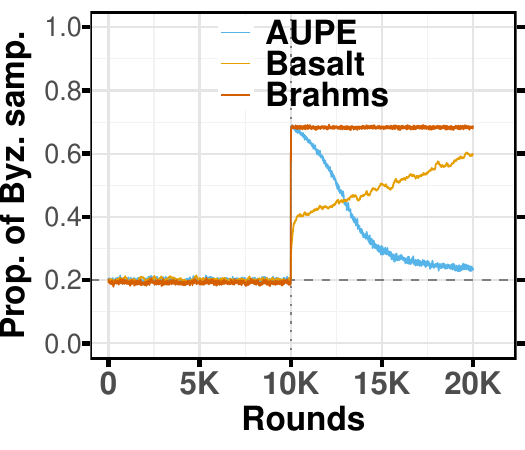}
    \caption{Full execution}
  \end{subfigure}%
  \begin{subfigure}{\mysubfigwidth}
    \includegraphics[width=\linewidth]{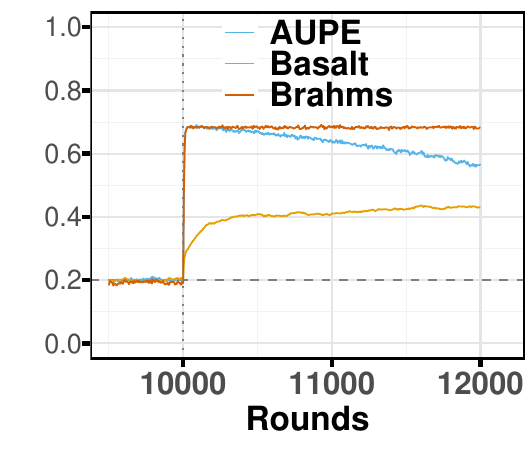}
    \caption{Attack onset}
  \end{subfigure}%
  \begin{subfigure}{\mysubfigwidth}
    \includegraphics[width=\linewidth]{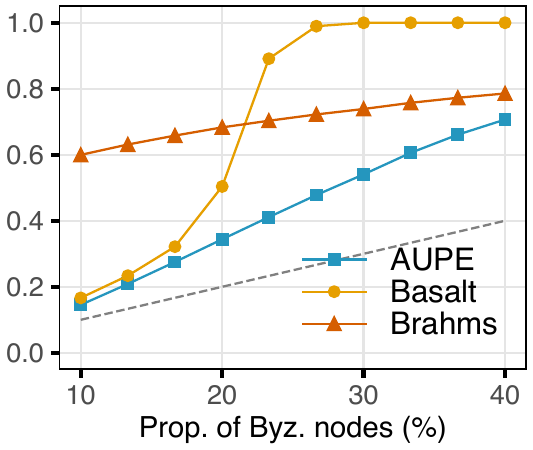}
    \caption{Average pollution}
  \end{subfigure}
  \caption{Delayed attack: execution and onset for $f=20\%$, and average pollution from rounds 11K to 20K.}
  \label{fig:delayed-baselines}
\end{figure}

To fill this gap, we present \emph{Hopper}, a bounded-memory debiasing peer-sampling protocol built on top of BRAHMS.
In each round, Hopper debiases the streams of received identifiers before supplying fresh candidates to the push and pull subviews.
Hopper relies on three building blocks:
(i) a bounded-memory occurrence estimator that preserves the distribution properties required for debiasing, (ii) a decay mechanism that keeps occurrence counts fresh over long executions, and (iii) a trusted collaboration mechanism that aggregates knowledge of overrepresented identifiers without relying on linear estimator mergeability.

We first select a bounded-memory occurrence estimator that preserves the distribution properties required for debiasing.
To do so, we compare bounded-memory estimators from the literature according to occurrence-distribution fidelity, separation of high- and low-frequency identifiers, and preservation of the adversarial bias factor. This study selects BitMatcher~\cite{bitmatcher:2024}, whose variable-width counters retain identifier fingerprints and allocate more capacity to frequent identifiers.
Second, we introduce a decay mechanism for BitMatcher, called \emph{BMDecay}, to keep occurrence counts fresh over long executions. At a capacity boundary, BMDecay extracts fingerprint-count pairs, reduces their counts, and reconstructs the estimator in decreasing count order.
This process removes stale occurrence counts while preserving the relative frequency gaps used for insertion during view construction.
Third, Hopper exploits trusted nodes provisioned with authentication secrets at bootstrap, allowing them to authenticate one another, merge their BMDecay estimators, and debias received streams using a dedicated trusted-node rule.
The trusted merge averages estimates only for bucket-scoped fingerprints present in both estimators. Entries present in only one estimator retain their original values, after which each estimator is reconstructed independently in decreasing-count order.
The resulting trusted BMDecay propagates evidence of identifiers that are repeatedly overrepresented across trusted observations rather than estimating the occurrence distribution observed by one node.
Trusted nodes therefore use a dedicated stream-debiasing rule that treats fingerprints identified as overrepresented differently from unknown fingerprints.

Our simulations show that Hopper-D substantially reduces view pollution under delayed attacks. Trusted-node collaboration further reduces peak pollution and provides parameter-dependent steady-state gains. This paper makes four contributions:
\begin{itemize}[leftmargin=*,nosep]
  \item \textbf{Delayed balanced attack}: We define an attack against cumulative-count debiasing and show why exact occurrence tracking does not ensure long-term resilience.
  \item \textbf{Adversarial estimator study}: We evaluate bounded-memory occurrence estimators on adversarial peer-sampling streams and select BitMatcher using debiasing-specific criteria.
  \item \textbf{Long-lived bounded-memory debiasing}: We design BMDecay, a capacity-triggered BitMatcher variant that preserves recent frequency gaps and fingerprint capacity over long executions.
  \item \textbf{Trusted collaboration}: We design fingerprint-aware merge and debiasing strategies that let trusted Hopper nodes aggregate evidence of overrepresented identifiers despite BitMatcher's non-linear state.
\end{itemize}

Section~\ref{sec:model} introduces the design foundations and system model. Section~\ref{sec:estimators} presents Hopper's architecture and estimator requirements, and selects the occurrence estimator that suits Hopper's objectives.
Section~\ref{sec:bmdecay} details Hopper, BMDecay, and trusted collaboration.
Section~\ref{sec:bmdecay-eval} evaluates their resilience and trusted-node leakage.
Section~\ref{sec:related-work} discusses related work, and Section~\ref{sec:conclusion} concludes.

\section{Background, system and adversarial model}
\label{sec:model}


\subsection{System Model}
\label{sec:system-model}


The system contains \(N\) active nodes: a fraction \(f\) of Byzantine nodes controlled by one adversary, a fraction \(t\) of trusted correct nodes, and a fraction \(h=1-f-t\) of ordinary correct nodes.
Trusted nodes are equipped with a TEE, such as Intel SGX or Intel TDX~\cite{sgx, cheng2024intel}, which provides code-integrity guarantees, remote attestation, and confidentiality for secret keys provisioned at bootstrap.
Each node possesses a unique identifier.
Nodes communicate over a routed network in loosely-synchronized logical rounds.

Each node maintains a view of \(v\) identifiers and periodically exchanges identifiers~\cite{peersampling:2004,jelasityGossipbased2007}.
We focus on steady membership, \eg, after bootstrap time \(T_0\), no node joins or leaves.
Hopper leverages BRAHMS's view-construction procedure.
Let \(\alpha\), \(\beta\), and \(\mu\) be the fractions assigned to the push, pull, and history subviews, respectively, with \(\alpha+\beta+\mu=1\).
At each round, a node sends its own identifier through push messages, requests views through pull messages, and obtains two candidate multisets: pushed identifiers and pulled identifiers from returned views.
It constructs the next view by drawing \(\alpha v\) entries from the push candidates, \(\beta v\) from the pull candidates, and \(\mu v\) from its history sample of size $l$.
The history subview is produced by min-wise sampling and converges to a Byzantine fraction bounded by \(f\) once sufficiently many correct identifiers have entered the sample stream.
The push and pull subviews, however, are drawn directly from fresh received identifiers and can therefore become biased toward Byzantine identifiers.




We rule out Sybil creation through the Sybil-resistance mechanisms assumed by BRAHMS, such as computational puzzles or equivalent rate limiting~\cite{Bortnikov:2008,douceurSybil2002}.
The adversary knows global membership and controls Byzantine protocol actions, but it does not know which nodes are trusted.
The adversary cannot break cryptographic primitives or corrupt trusted or correct code.
TEE side channels, rollback, and attestation-root compromise are outside our model.

\subsection{Debiasing in AUPE}

AUPE~\cite{aupe:2024} adds a Set Cleanser before the push and pull subviews are built for the final view $v$.
For each node \(u\), let \(\sigma_u^r\) be the multiset of identifiers received by \(u\) in round \(r\), and let
\[
  \Phi_u^r(x) = \sum_{\tau \le r} \mathrm{occ}_{\sigma_u^\tau}(x)
\]
be the cumulative number of times \(u\) has received identifier \(x\) up to round \(r\).
AUPE stores \(\Phi_u^r\) in a tracking component, typically an array.
For each identifier \(x\) from a stream received in round \(r\), AUPE inserts \(x\) into a small local sample memory with probability
\[
  p_u^r(x) = \frac{\text{min}_u^r}{\Phi_u^r(x)},
\]
where $\text{min}_u^r$ is the minimum non-zero value stored in $\Phi_u^r$.
Thus, identifiers that have appeared frequently are sampled less often.
For each received identifier, one identifier is drawn uniformly at random from the latter sample memory to fill an output stream from which $\alpha v$ and $\beta v$ identifiers are drawn to construct the push and pull subviews of the final view.
This debiasing strategy is applied every round, so its output also influences future pull replies.





\subsection{Adversarial stream modeling}
\label{sec:attack-model}

We consider the balanced overrepresentation attack~\cite{Bortnikov:2008}.
After \(T_a\), coordinated Byzantine nodes advertise Byzantine identifiers through pushes and pull replies, maximizing system-wide propagation while respecting BRAHMS's push limits.

Let \(E_B\) and \(E_C\) be the sets of Byzantine and non-Byzantine identifiers, respectively.
Within a window \(W\) after \(T_a\), a non-Byzantine node receives a stream containing a proportion \(w\) of Byzantine identifiers and \(1-w\) of non-Byzantine identifiers.
Under a balanced attack, identifiers are approximately uniform within each identifier class. We define the class bias factor as
\[
  \gamma =
  \frac{w/|E_B|}{(1-w)/|E_C|}.
\]
When \(\gamma=1\), Byzantine and correct identifiers have the same expected per-identifier frequency.
When \(\gamma>1\), each Byzantine identifier appears \(\gamma\) times more often than each correct identifier in the stream.

\subsection{Delayed Balanced Attack}
\label{sec:delayed-attack}

The delayed balanced attack exploits the freshness weakness of AUPE's cumulative exact counting.
Before \(T_a\), Byzantine nodes behave similarly to correct nodes. Consequently, the exact occurrence counts maintained for Byzantine and non-Byzantine identifiers are both approximately \(c\), and their insertion probabilities are similar.

After \(T_a\), the adversary launches a balanced attack with live bias factor \(\gamma>1\).
If a non-Byzantine identifier has expected frequency \(\lambda\) in the received stream at round \(r\), a Byzantine identifier has expected frequency \(\gamma\lambda\).
After \(\Delta\) attack rounds, the expected cumulative ratio observed by AUPE between Byzantine and non-Byzantine identifiers is roughly
\[
  \frac{c+\gamma\lambda\Delta}{c+\lambda\Delta}.
\]
When \(c \gg \lambda\Delta\), this ratio remains close to one even if \(\gamma\) is large.
AUPE therefore assigns Byzantine and non-Byzantine identifiers nearly equal insertion probabilities even though the input stream contains many more Byzantine occurrences. Byzantine identifiers consequently remain overrepresented in the output stream and in correct nodes' views.
%
This attack demonstrates the failure of weighting all historical occurrences equally and motivates Hopper.



\section{Hopper Estimator}
\label{sec:estimators}

Hopper uses a bounded-memory estimator to reduce the insertion probability of overrepresented identifiers into the sample memory used to construct each node's push and pull subviews. This section defines the estimator requirements, evaluates bounded-memory candidates, analyzes the mergeability trade-off, and describes the selected estimator, BitMatcher.



\subsection{Debiasing-Relevant Properties Under Adversarial Streams}
\label{sec:estimator-properties}

Although a fixed-size estimator necessarily sacrifices some precision, an estimator suitable for stream debiasing must preserve the occurrence-distribution properties that drive Hopper's insertion probability.

Overestimating a correct identifier suppresses a potentially useful insertion into the sample memory, whereas underestimating a Byzantine identifier inserts it too often.
Because correct identifiers form the larger class in the considered settings, average error can hide poor estimates for the smaller Byzantine class.
We therefore use three complementary properties relevant to the debiasing of identifier streams under adversarial conditions.
We then study the freshness of the information stored and the mergeability of the selected estimator in Section~\ref{sec:bmdecay-eval}.

\textbf{Distribution fidelity.}
Distribution fidelity quantifies agreement between the complete exact and estimated occurrence distributions. For normalized exact and estimated count vectors $p$ and $q$, respectively, it is $D_{\mathrm{KL}}(p\Vert q)$. Lower values are better, but do not guarantee that the smaller Byzantine class remains distinguishable.

\textbf{Class separability.}
Class separability quantifies whether the estimator preserves the high- and low-frequency classes. The $F_1$-score compares $K=2$ clusters of estimated counts with the ground-truth overrepresented and underrepresented classes. A value near one indicates that collisions and replacement have not erased the frequency gap. Clustering is used only for evaluation as Hopper performs no classification during its execution.

\textbf{Bias-factor preservation.}
Bias-factor preservation measures the relative occurrence-frequency signal that drives insertion into Hopper's sample memory. Let $\gamma$ be the ratio of the overrepresented to underrepresented class-average exact counts, and $\hat{\gamma}$ the analogous ratio of estimated counts.
Their relative error is $\gamma_{\mathrm{err}}=(\hat{\gamma}-\gamma)/\gamma$.
Zero is exact, a negative value implies insufficient suppression of overrepresented identifiers, whereas a positive value implies excessive suppression.

Effective debiasing therefore requires low distribution divergence, high class-separation $F_1$, and bias-factor error close to zero.




\subsection{Estimator evaluation method}
\label{sec:estimator-eval}

We compare estimators that provide bounded \(B\)-byte state and inexpensive updates and queries: Count-Min Sketch with Conservative Updates (CMSCU), Count-Mean-Min with Conservative Updates (CMMCU), Lossy Conservative Update (LCU)~\cite{cormode:2005,estan:2003,deng:2007new,goyal:2011,goyal:2012}, Cold Filter (CF)~\cite{coldfilter:2019}, XY~\cite{liu:2023}, and BitMatcher (BM)~\cite{bitmatcher:2024}.

Table~\ref{tab:sketches} summarizes their representation-specific strengths and risks. We then vary memory budget, stream length, and population size to stress the properties required for peer-sampling debiasing.

\begin{table}[!t]
    \centering
    \caption{Candidate estimators for Hopper's debiasing}
    \label{tab:sketches}
    \footnotesize
    \setlength{\tabcolsep}{2pt}
    \renewcommand{\arraystretch}{0.95}
    \resizebox{\columnwidth}{!}{%
        \begin{tabular}{P{0.10\textwidth} P{0.16\textwidth} P{0.2\textwidth}}
            \toprule
            \textbf{Estimator}                             & \textbf{Strength}                                                   & \textbf{Risk for Hopper}                                                 \\
            \midrule
            CMSCU~\cite{estan:2003}                        & Simple, mergeable, stable in uniform streams                        & Collision overestimation. Hides high/low-frequency gap                   \\
            \midrule
            CMMCU~\cite{goyal:2012}, LCU~\cite{goyal:2011} & Reduce Count-Min overestimation                                     & Underestimates high-frequency IDs                                        \\
            \midrule
            Cold Filter~\cite{coldfilter:2019}             & Separates cold and hot IDs                                          & Threshold tuning is workload dependent                                   \\
            \midrule
            XY~\cite{liu:2023}                             & Compact probabilistic estimator                                     & Poor class preservation under strong adversarial bias                    \\
            \midrule
            BitMatcher~\cite{bitmatcher:2024}              & Adapts counter widths to skew, preserves hot IDs under tight memory & Non-linear, fingerprint-coupled, and not directly mergeable or age-aware \\
            \bottomrule
        \end{tabular}%
    }
\end{table}

\begin{figure}[!t]
  \centering
  \includegraphics[width=.75\linewidth]{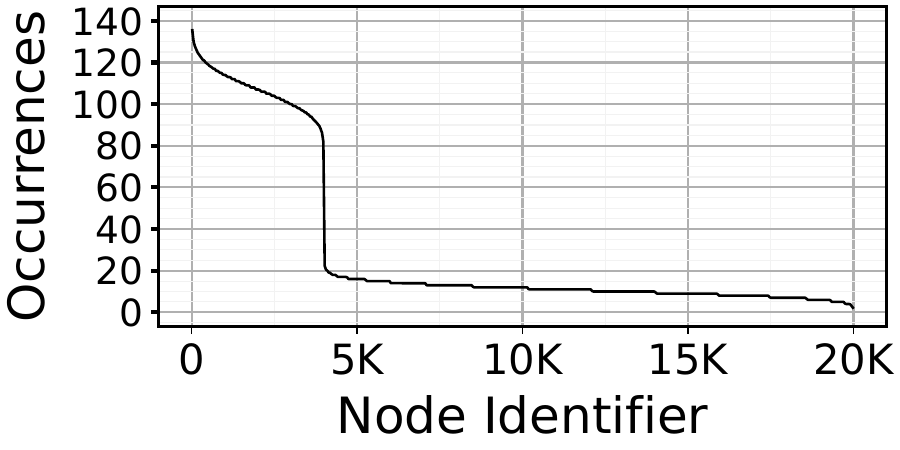}
  \caption{Identifier distribution (\(N=20\)K, \(M=600\)K, \(f=20\%\), \(\gamma=10\)).}
  \label{fig:input-streams}
\end{figure}

We generate streams inspired by Bitcoin peer-discovery scale: \(N=20\)K distinct identifiers and \(M=600\)K to 7.2 million observations, corresponding to approximately 30 days through one year at 20K received identifiers per day~\cite{bitnode:07072025}. An exact array of four-byte counters requires 80 KB, estimator budgets range from 20 to 80 KB.
Adversarial streams use \(f\in\{10\%,20\%,30\%\}\) and bias factor \(\gamma=10\). Figure~\ref{fig:input-streams} shows one resulting occurrence distribution.
Every Byzantine identifier receives weight \(\gamma\), every correct identifier weighs one, and identifiers are drawn from the normalized population.
Count-Min variants use three hash functions. Cold Filter assigns 90\% of memory to filtering layers and 10\% to its CMSCU backing sketch~\cite{goyal:2012,coldfilter:2019}.

\begin{figure}[!t]
  \centering
  \includegraphics[width=\linewidth]{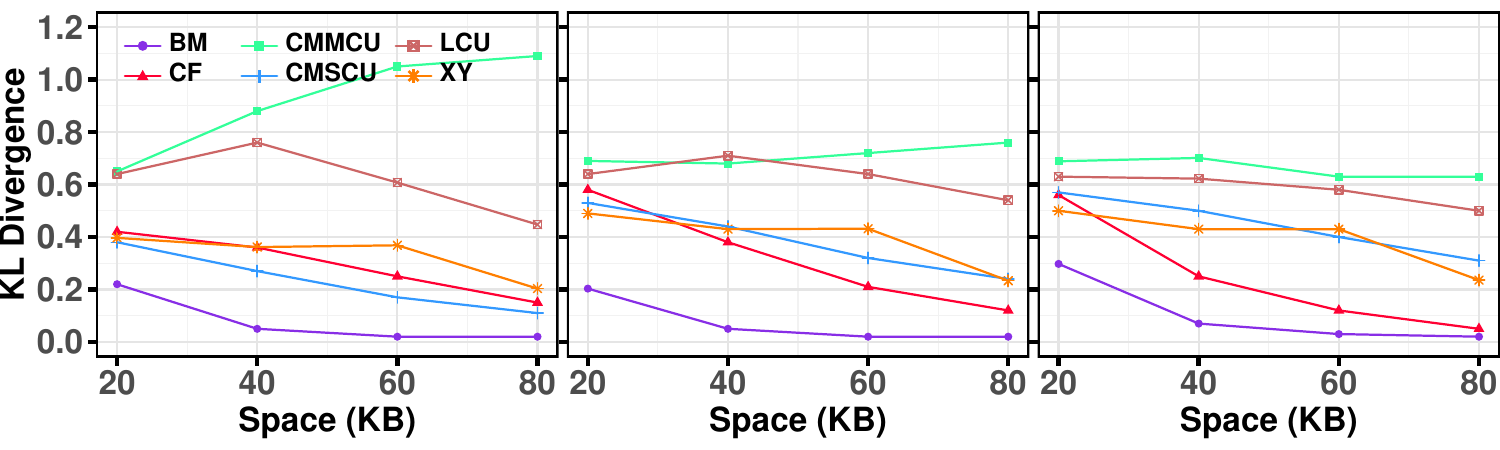}
  \includegraphics[width=\linewidth]{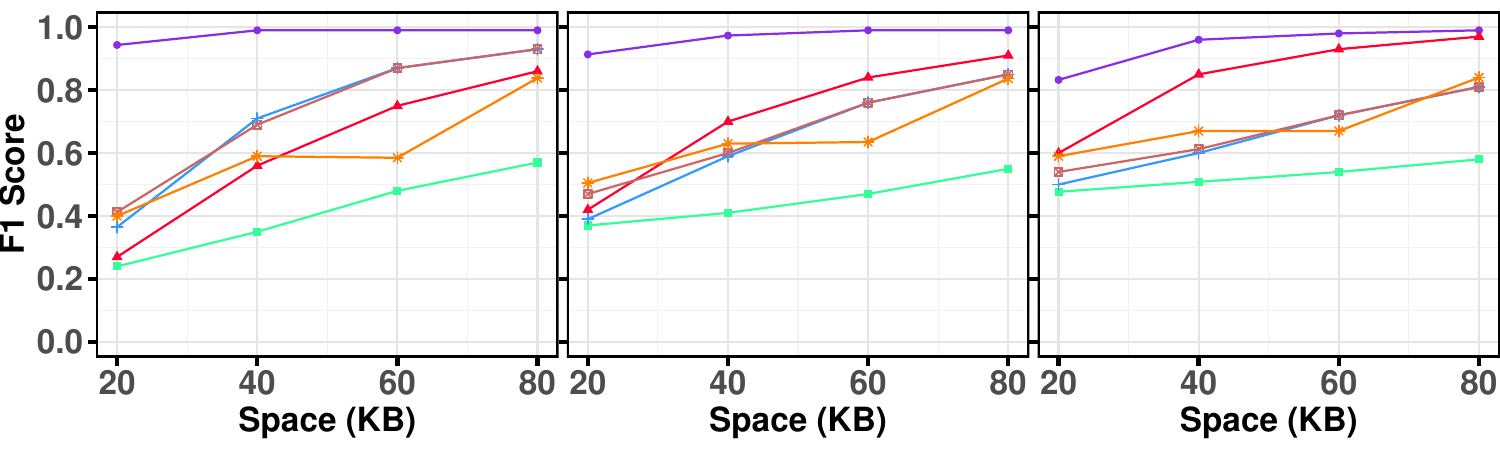}
  \includegraphics[width=\linewidth]{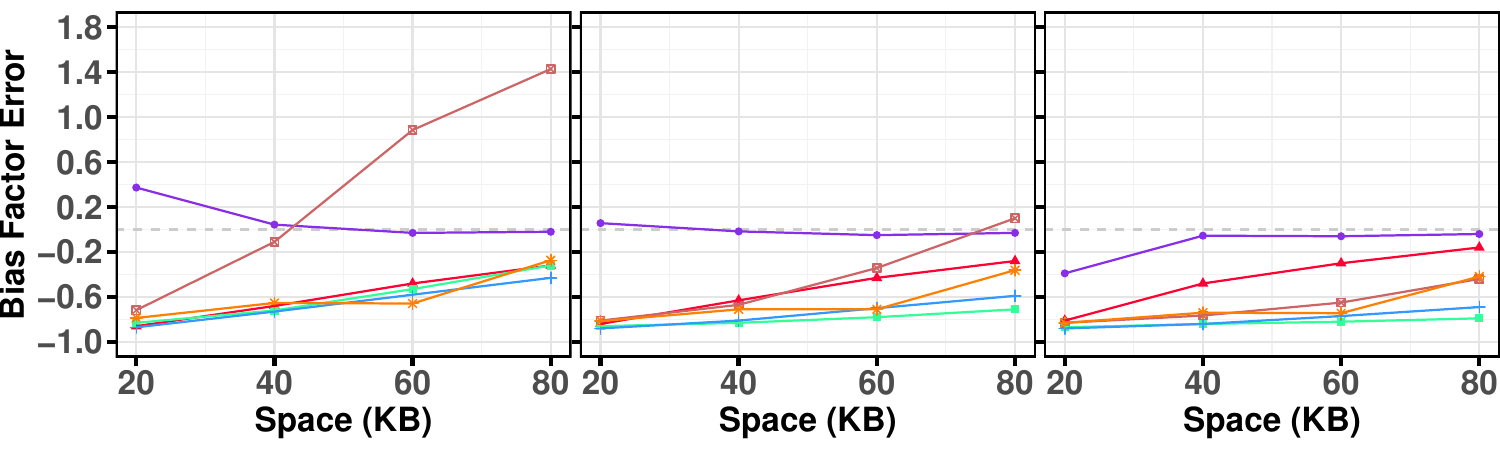}
  \caption{Estimator quality versus memory: KL divergence, \(F_1\), and bias-factor error. Columns show \(f=10\%,20\%,30\%\).}
  \label{fig:sketch-budget}
\end{figure}

\begin{figure}[!t]
  \centering
  \includegraphics[width=\linewidth]{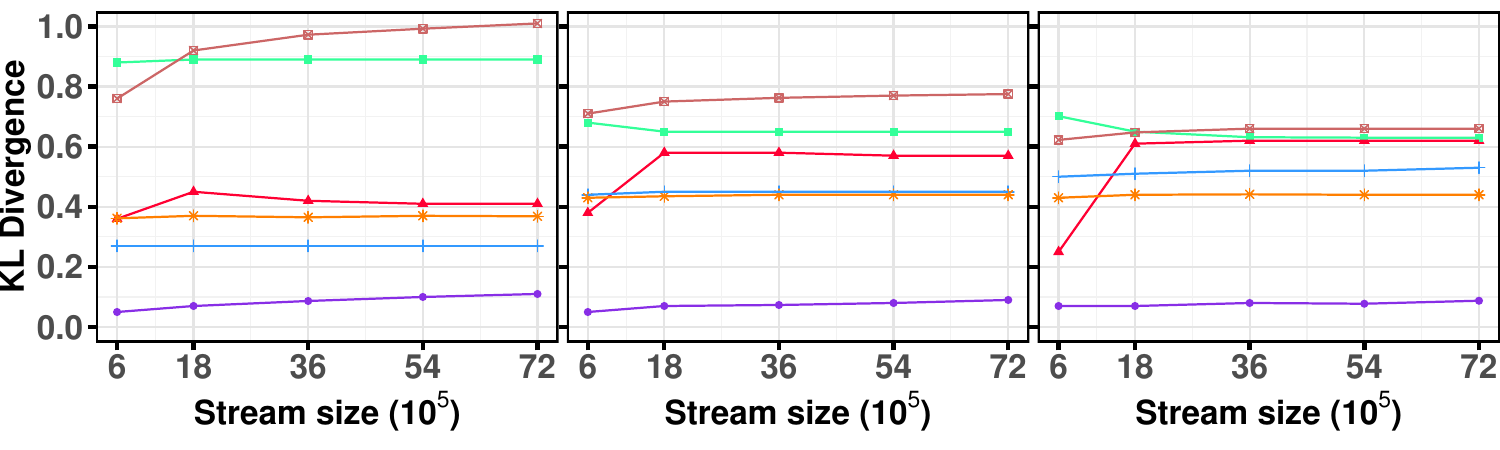}
  \includegraphics[width=\linewidth]{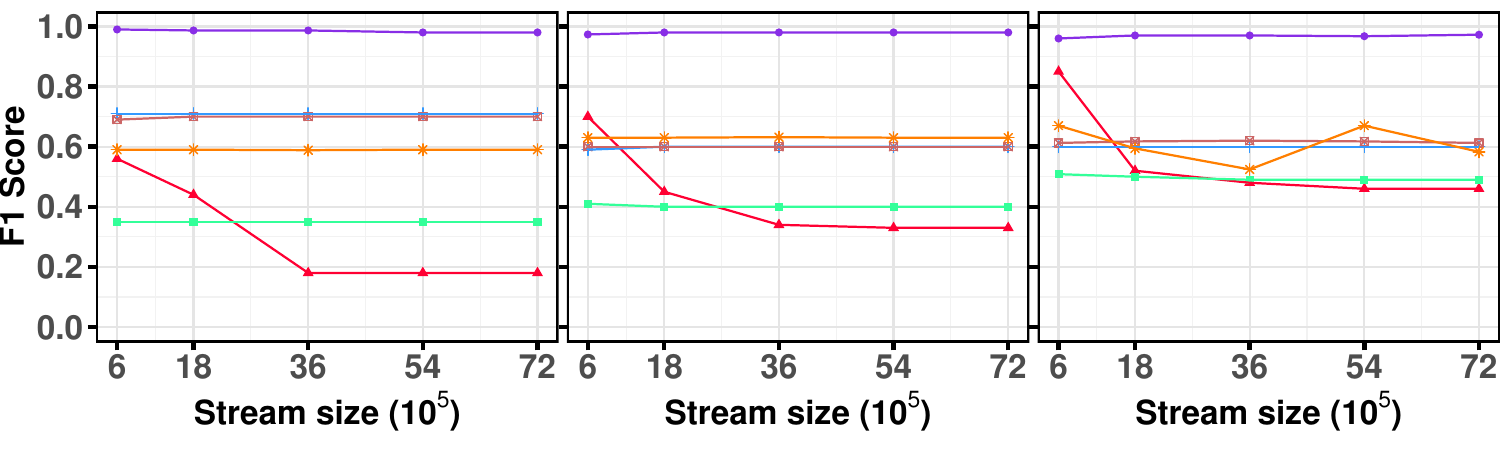}
  \includegraphics[width=\linewidth]{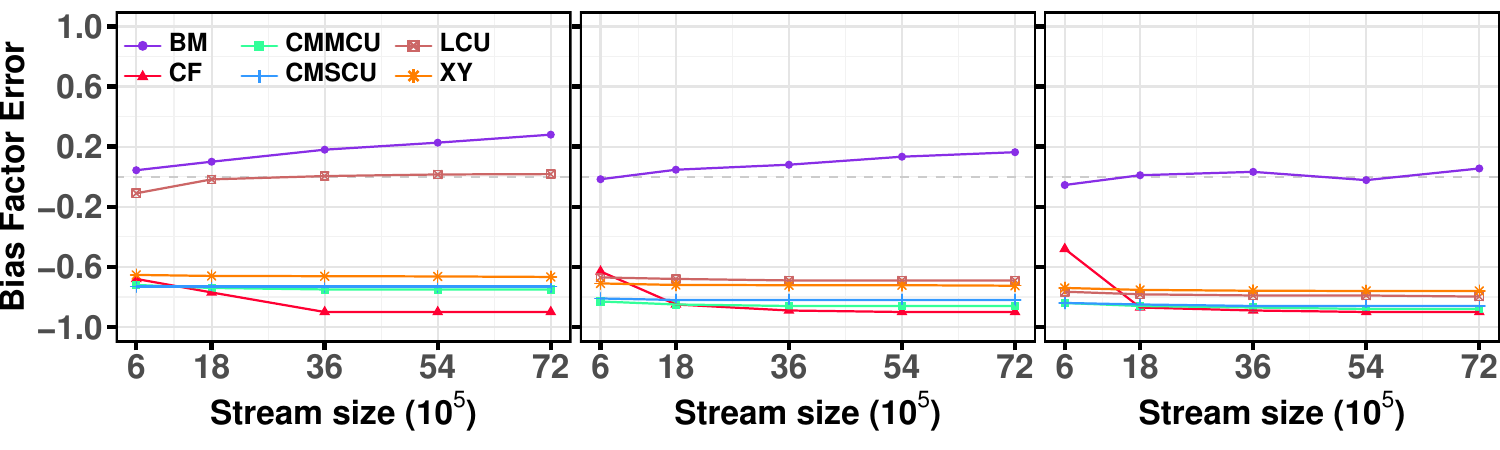}
  \caption{Estimator quality versus stream length at 40 KB. Columns show \(f=10\%,20\%,30\%\).}
  \label{fig:sketch-stream}
\end{figure}

\begin{figure}[!t]
  \centering
  \includegraphics[width=\linewidth]{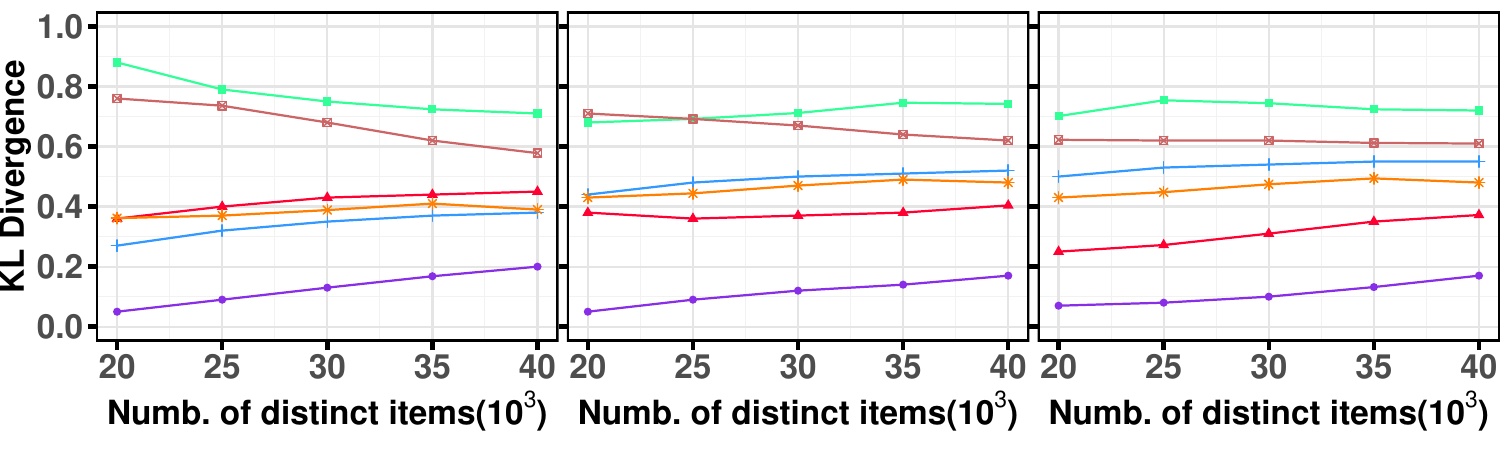}
  \includegraphics[width=\linewidth]{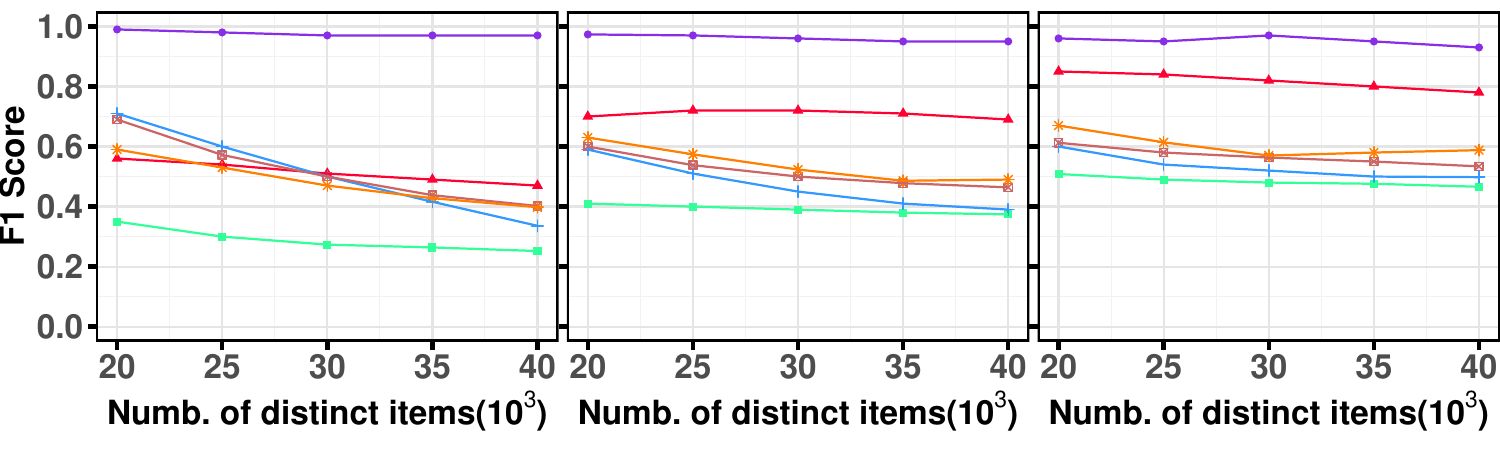}
  \includegraphics[width=\linewidth]{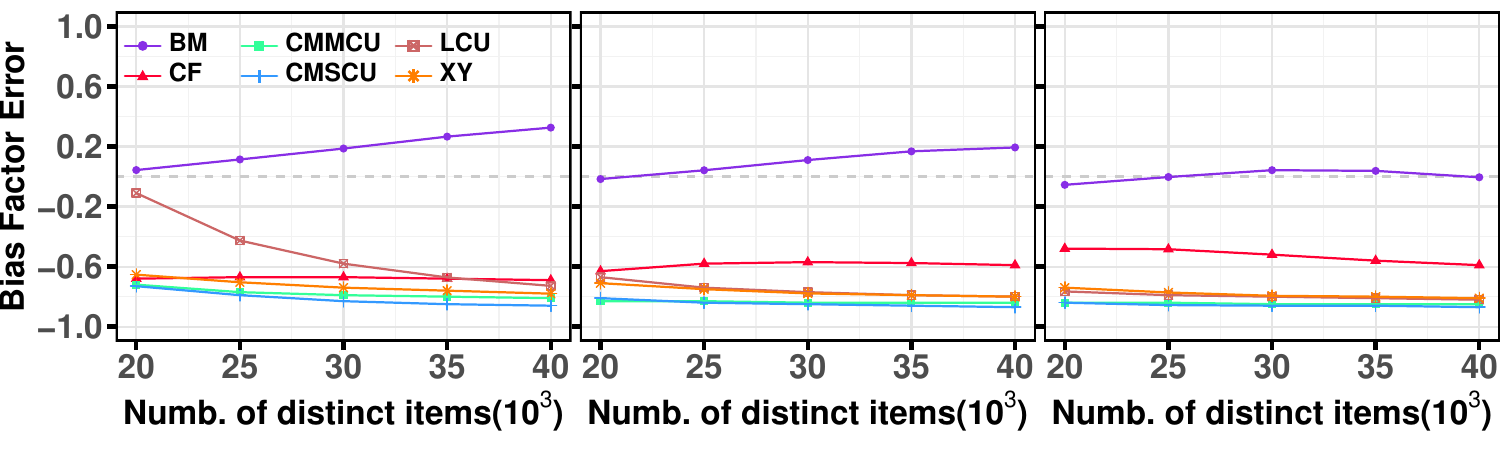}
  \caption{Estimator quality versus population size at 40 KB. Columns show \(f=10\%,20\%,30\%\).}
  \label{fig:sketch-n}
\end{figure}

Each microbenchmark ingests one generated stream, queries every identifier after ingestion, and reports KL divergence, $F_1$-score, and bias-factor error. Figures~\ref{fig:sketch-budget}, \ref{fig:sketch-stream}, and \ref{fig:sketch-n} vary memory budget, stream length, and population size, respectively.


\subsection{Estimator evaluation results}
\label{sec:adversarial-sketch-results}

Across the evaluated settings, BM provides the best joint combination of low distribution divergence, high class separability, and bias-factor error close to zero. Its estimates nevertheless become stale as the stream grows, as reflected by increasing bias-factor error. We detail both observations below.

\subsubsection{Memory pressure}
BM provides the best joint result across all Byzantine fractions. It has the lowest KL divergence and the highest $F_1$-score, including at the 20 KB budget.
Its bias-factor error is not always zero under this tightest budget, but becomes close to zero from 40 KB onward.
The alternatives improve with memory but do not preserve all three properties simultaneously. Most of them underestimate $\gamma$.

\subsubsection{Stream-length pressure}
BM keeps $F_1$ close to one and KL divergence below the alternatives throughout.
However, its bias-factor error drifts upward for $f=10\%$ and $20\%$, even though class separability remains high.
BM retains the two frequency classes but progressively distorts their relative magnitude, exposing the lack of an aging mechanism for unbounded streams.

\subsubsection{Population pressure}
Increasing $N$ from 20K to 40K at fixed memory similarly increases BM's KL divergence and its bias-factor error for $f=10\%$ and $20\%$, while its $F_1$ remains above 0.9.
The alternatives generally lose class separability and continue to underestimate the adversarial bias.

\subsection{Mergeability and selection trade-off}
\label{sec:selection-tradeoff}

Mergeability matters because trusted Hopper nodes combine evidence collected from different streams. A mergeable summary can represent the multiset union of two input streams without replaying either stream~\cite{cmsmerge}. In a standard Count-Min sketch, identically configured nodes assign the same meaning to corresponding counters, cell-wise addition is therefore commutative, associative, and equivalent to processing both streams in one sketch.

BitMatcher does not have these linear semantics. Its stream-adaptive layout may place the same fingerprint in different arrays, bucket states, or counter widths at two nodes, while equal memory positions may hold unrelated fingerprints. Position-wise addition is therefore invalid, and a short fingerprint alone is ambiguous across bucket neighborhoods. Hopper instead extracts entries as \((fp,h_1,c)\), matches them by the composite key \((fp,h_1)\), averages counts only for keys present in both estimators, and independently reconstructs each bounded state in decreasing-count order (Section~\ref{sec:trusted-merge}). This operation is lossy and non-associative rather than a linear stream-union summary.

Mergeability is useful for collaboration, but cannot recover a frequency gap already erased by estimation error. Because local debiasing occurs every round, Hopper prioritizes preservation of the adversarial frequency signal and selects BitMatcher despite the need for custom reconstruction.

\subsection{BitMatcher}
\label{sec:why-bitmatcher}
In this section, we summarize BitMatcher's design and operations, which are detailed in~\cite{bitmatcher:2024}.
BitMatcher adapts fixed memory to skewed streams~\cite{bitmatcher:2024}: cold identifiers use small counters, while overflows reallocate bits toward hot identifiers. Each bucket therefore adapts to its local frequency context.

\textbf{Structure.}
BitMatcher maintains two arrays $A_1$ and $A_2$ of fixed-size buckets. Each bucket contains entries that couple a fixed-size fingerprint with a variable-width counter, a state flag specifies the current bit allocation. Hopper initially divides a 64-bit bucket into five 8-bit fingerprints paired with counters of 2, 3, 4, 5, and 6 bits, plus a 4-bit state flag. Later states may retain four or fewer fingerprints and enlarge the hottest counters.
BitMatcher thus trades distinct-fingerprint capacity for counter magnitude.

Identifier $x$ has fingerprint $fp(x)$ and two candidate buckets, one in each array. The first bucket is $h_1(x)=H(x)$, the alternate bucket is derived from the first bucket and the fingerprint:
\[
  h_1(x)=H(x), \qquad h_2(x)=h_1(x)\oplus H(fp(x)).
\]
This partial-key cuckoo construction recovers an alternate bucket from the current bucket and fingerprint, without storing the identifier.

\textbf{Insertion.}
BitMatcher scans both candidate buckets. It increments a matching fingerprint or stores a new one with count one in an empty entry. An overflowing entry may first exchange places with a colder entry in a larger counter.

When both buckets are full, BitMatcher decrements the smallest-width entry in one candidate bucket and replaces its fingerprint only when its count reaches zero. This protects accumulated evidence while eventually admitting new identifiers.

If no larger slot can absorb an overflow, BitMatcher changes the bucket state. When the largest counter overflows, it removes the smallest entry and assigns its bits to the largest counter, storing one fewer fingerprint. When a smaller counter overflows, it first shrinks the largest counter if its value fits in fewer bits and redistributes the released bits. Otherwise, it attempts bounded cuckoo relocation or removes the smallest entry. This is \emph{bit-level matching}: observed overflows change individual counter widths rather than a fixed allocation.

\textbf{Query.}
BitMatcher searches both buckets for $fp(x)$. A match returns its counter, an absent fingerprint returns zero if a bucket has an empty entry, or the minimum candidate counter if both are full. Collisions and replacement introduce error, but retained fingerprints distinguish represented identifiers from unknown ones.

\section{Hopper Design}
\label{sec:bmdecay}

\subsection{Per-Round Operation and Variants}
\label{sec:hopper-operation}

Hopper retains BRAHMS's view construction but adds a bounded-memory debiasing step to the received streams of pushed and pulled identifiers before directly sampling the push and pull subviews of the final node view $v$.
In round \(r\), node \(u\) receives multiset \(\sigma_u^r\).
For each identifier $x$ in the input stream, the node inserts $x$ into estimator \(S_u\) and queries its estimated occurrence count.
Let \(\hat{\Phi}_u(x)\) be its estimated count and \(m_u\) the minimum non-zero count retained by \(S_u\).
The node then inserts \(x\) into its sampling memory with probability
\[
  p_u(x)=\frac{m_u}{\hat{\Phi}_u(x)}.
\]
Hopper processes every pushed and pulled identifier sequentially with the same probability, drawing an identifier uniformly from the sample memory after each update to produce separate debiased output push and pull streams. It samples \(\alpha v\) and \(\beta v\) identifiers from these streams to construct the push and pull subviews, respectively, and retains BRAHMS's \(\mu v\)-identifier min-wise history subview.
Each repeated occurrence increases the identifier's estimate and triggers another insertion trial using the updated probability.
The sample memory persists across rounds.

\subsection{Hopper in Indefinitely Living Systems with BMDecay}
\label{sec:hopper-decay}

BitMatcher provides a fixed memory footprint, but two limitations prevent its direct use in an indefinitely living system.
First, BitMatcher accumulates occurrences without aging them. As the stream grows, historical observations dominate its counters and new observations have diminishing influence. The estimated ratio between overrepresented and underrepresented identifiers can therefore lag behind the current stream.
Its bias-factor error drifts even while class separability remains high (Section~\ref{sec:adversarial-sketch-results}). Because Hopper's insertion rule depends on this ratio, preserving the two classes is insufficient as a long benign stream can mask the frequency change caused by a delayed attack.
%
Second, BitMatcher handles counter overflow by changing a bucket's state and reallocating bits from fingerprint--counter entries to larger counters. Some transitions remove entries, under sustained insertions the estimator still occupies \(B\) bytes but retains fewer distinct fingerprints. New fingerprints are then increasingly rejected or replace existing entries, degrading occurrence estimation and class separability.

\subsubsection{Capacity-triggered decay}
BMDecay is a decay-enabled BitMatcher variant that preserves relative frequency gaps while retaining at least four fingerprint--counter entries per 64-bit bucket. Starting from counter widths \(\langle2,3,4,5,6\rangle\), any transition that would leave fewer than four entries triggers estimator-wide decay instead of removing another entry. This capacity boundary prevents historical counters from indefinitely consuming fingerprint slots.


\subsubsection{Decay procedure}

Upon decay, BMDecay extracts all \((fp,h_1,c)\) entries, replaces each count \(c\) by \(\lfloor c/2\rfloor\), discards zeros, sorts survivors by decreasing count, clears the estimator, and reinserts the survivors in that order. For entries stored in the second array, it recovers the canonical bucket as \(h_1=h_2\oplus H(fp)\). The composite key \((fp,h_1)\) preserves the bucket neighborhood, while sorting prioritizes fingerprints with larger counts.
For entries whose halved counts remain non-zero, the decay procedure approximately preserves the multiplicative contrast:
\[
  \frac{\lfloor c_x/2\rfloor}{\lfloor c_y/2\rfloor}\approx\frac{c_x}{c_y}.
\]
Repeated decay geometrically reduces history so that new observations matter again, while discarded entries restore fingerprint capacity.
BMDecay is event-driven and provides neither cumulative nor sliding-window semantics.
For \(C\) retained entries, decay costs \(O(C\log C)\) operations, and occurs only at the state boundary.

\subsection{Trusted Merge}
\label{sec:trusted-merge}

In this section, we describe how trusted nodes can combine their BMDecay estimators to improve debiasing despite the non-linear nature of BitMatcher.

As in RAPTEE~\cite{Pigaglio:2022}, Hopper nodes invoke a TEE-backed authentication protocol before estimator exchange. All nodes invoke the protocol to avoid explicitly disclosing their role, but only trusted nodes possessing the provisioned secret complete authentication and exchange estimators.

BitMatcher entries cannot be merged position-wise because equal fingerprints may occupy different arrays and bucket states.
Algorithm~\ref{alg:trusted-merge} first converts each entry to a canonical triple \((fp,h_1,c)\). For an entry stored in the second array, canonical extraction recovers \(h_1=h_2\oplus H(fp)\). Within one estimator, BitMatcher maintains at most one entry for a bucket-scoped fingerprint \((fp,h_1)\), distinct identifiers that collide on this key remain indistinguishable.

The merge modifies only bucket-scoped fingerprints present in both estimators. Their two occurrence estimates are replaced by their average in both outputs. A fingerprint present in only one estimator retains its original estimate in that estimator and remains absent from the other. Consequently, the two reconstructed estimators may contain different fingerprint sets.

\begin{algorithm}[!b]
  \caption{Hopper trusted merge}
  \label{alg:trusted-merge}
  \KwIn{Authenticated snapshots \(S_u,S_v\), budget \(B\)}
  \KwOut{Replacement estimators \(S'_u,S'_v\)}
  \(U \gets \operatorname{CanonicalExtract}(S_u)\)\;
  \(V \gets \operatorname{CanonicalExtract}(S_v)\)\;
  \(\mathcal{I}_u \gets \operatorname{Entries}(U)\); \(\mathcal{I}_v \gets \operatorname{Entries}(V)\)\;
  \ForEach{\(k \in \operatorname{Keys}(U)\cap\operatorname{Keys}(V)\)}{
  \(c \gets \lfloor(U[k]+V[k])/2\rfloor\)\;
  \(\mathcal{I}_u[k].c \gets c\); \(\mathcal{I}_v[k].c \gets c\)\;
  }
  Sort each \(\mathcal{I}_i\) by decreasing \(c\), then increasing \((h_1,fp)\)\;
  \(S'_u \gets \operatorname{DeterministicRebuild}(\mathcal{I}_u,B)\)\;
  \(S'_v \gets \operatorname{DeterministicRebuild}(\mathcal{I}_v,B)\)\;
  \Return \(S'_u,S'_v\)\;
\end{algorithm}

Each output list is sorted independently because it contains the fingerprints originally retained by its corresponding estimator. \(\operatorname{DeterministicRebuild}\) initializes counters directly from the ordered canonical triples, it does not replay \(c\) insertions or recursively trigger decay. It uses deterministic bucket placement and bounded relocation, and discards an entry if neither candidate bucket can retain it within budget \(B\).

Exchanging the inputs exchanges the outputs. When both inputs contain the same entries and reconstruction retains them all, it is logically idempotent. It is neither linear nor generally associative because averaging, flooring, and bounded reconstruction lose information. Additionally, it does not compute the sum or union of the input streams, instead it shares occurrence information only for fingerprints represented by both estimators.

For at most \(C\) extracted entries and bounded relocation depth, merge costs \(O(C\log C)\), communicates \(O(B)\) bytes, and remains exposed to collisions of the bucket-scoped fingerprint \((fp,h_1)\).

\subsection{Trusted Debiasing}
\label{sec:trusted-debiasing}

After merge, let \(c_{\min}\) and \(c_{\max}\) be the minimum and maximum positive retained counts.
Because trusted merge preferentially retains high-count fingerprints, the presence of a candidate's fingerprint is treated as evidence of broad overrepresentation across trusted observations. Trusted nodes therefore use separate insertion probabilities for known and unknown fingerprints:
\[
  p_{\mathrm{known}}=\frac{c_{\min}}{c_{\max}},
  \qquad
  p_{\mathrm{unknown}}=\frac{1}{c_{\min}}
\]
The first probability applies when the fingerprint of \(x\) is present and the second otherwise. This role-specific rule is a heuristic, not a Byzantine classifier.

\paragraph*{Security Properties of Collaboration}
\label{sec:merge-security}

The mutual authentication protects the trusted merge operations.
An unauthenticated node cannot submit an arbitrary estimator for reconstruction inside a trusted node.
The TEE protects merge code and authentication secrets, but Byzantine identifiers can still be inserted in the estimator through normal push and pull messages.
Trusted merging can nevertheless produce statistically distinguishable views. Section~\ref{sec:bmdecay-eval} evaluates a view-based trusted-node inference attack to assess this risk.

\section{Hopper Evaluation}
\label{sec:bmdecay-eval}

We ask four questions: Does decay preserve long-lived estimates (RQ1)? Does decay improve resilience to the delayed balanced attack (RQ2)? What benefit does trusted collaboration provide (RQ3)? Does collaboration reveal trusted nodes to the adversary (RQ4)?

\subsection{Implementation and Setup}
\label{sec:bmdecay-setup}

We integrated the C++ BitMatcher implementation~\cite{bitmatcher_code} into our Rust Hopper simulator. Unless stated otherwise, protocol experiments use \(N=1{,}000\), \(v=20\), \(f\in\{10\%,20\%,30\%,40\%\}\), and a 500-byte budget for BM or BMDecay, equal to 12.5\% of AUPE's four-byte exact-counter array. Byzantine nodes behave correctly until round \(T_a=10{,}000\), then launch the attack from Section~\ref{sec:delayed-attack}. We used a sample memory size of 10 identifiers.

We report Byzantine-view pollution averaged over non-Byzantine nodes. Uniform sampling has target \(f\). AUPE uses exact cumulative counters, Hopper uses BM, and Hopper-D differs only by using BMDecay. BRAHMS and BASALT are non-debiasing baselines. Protocol comparisons share membership, attack schedule, and view construction. RQ1 is a single-estimator microbenchmark using the metrics from Section~\ref{sec:estimator-properties}. Trusted collaboration is enabled only in RQ3 and RQ4.



\subsection{RQ1: Does Decay Preserve Long-Lived Estimates?}

\begin{figure}[!b]
  \centering
  \includegraphics[width=\linewidth]{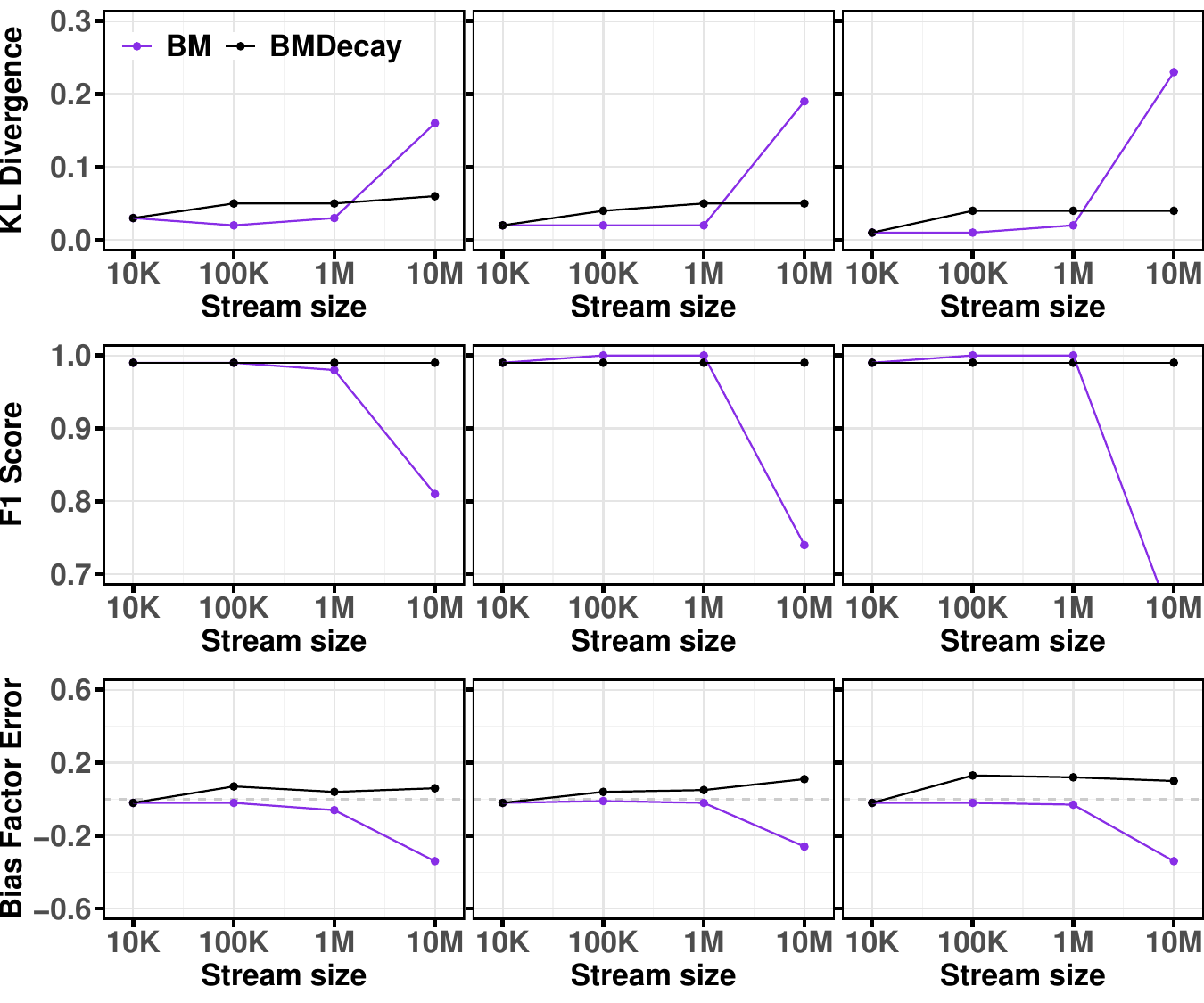}
  \caption{BM and BMDecay over long streams: KL divergence, \(F_1\), and bias-factor error
  }
  \label{fig:bmdecay-stream-quality}
\end{figure}

\begin{figure}[!b]
  \centering
  \includegraphics[width=\linewidth]{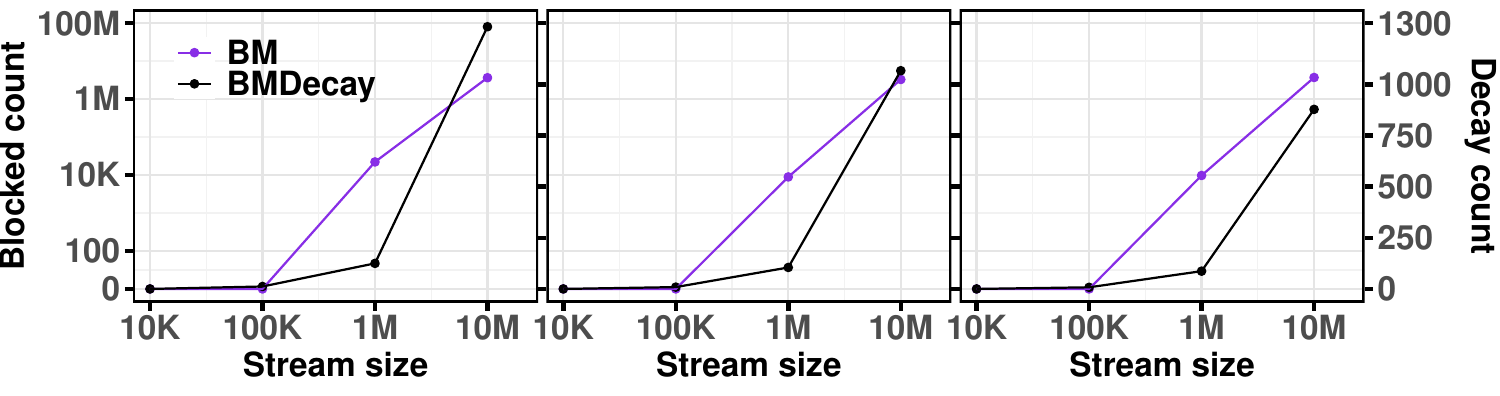}
  \caption{Blocked BM insertions (left axis) and BMDecay events (right axis). The columns show \(f=10\%,20\%,30\%\).}
  \label{fig:blocked-decay}
\end{figure}

We feed BM and BMDecay streams of up to \(10^7\) identifiers with $\gamma=10$ and \(f\in\{10\%,20\%,30\%\}\). Figure~\ref{fig:bmdecay-stream-quality} shows similar quality through \(10^6\) insertions. At \(10^7\), BM's KL divergence rises, its bias-factor error becomes negative, and its \(F_1\) drops by up to about 40 percentage points. BMDecay keeps \(F_1\) near one and bias-factor error bounded, with a small positive error.
Figure~\ref{fig:blocked-decay} reports BM insertions blocked because an overflowing bucket cannot transition to a state with fewer entries. By \(10^7\) identifiers, BM blocks millions of insertions, whereas BMDecay converts saturation into roughly \(10^3\) decay procedures.
Decay therefore preserves both frequency information and fingerprint capacity.

\subsection{RQ2: Does Decay Improve Protocol Resilience?}

\begin{figure}[!t]
  \centering
  \includegraphics[width=\linewidth]{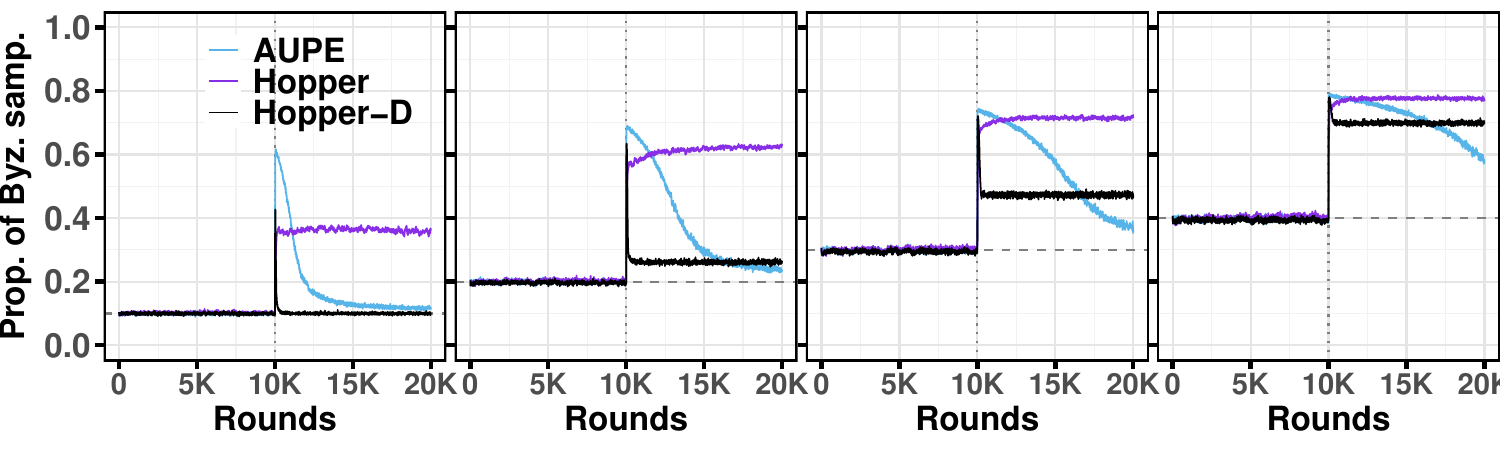}
  \caption{View pollution under the delayed attack. The vertical line marks $T_a$=10K. Columns show $f$=10\%,20\%,30\%,40\%.
  }
  \label{fig:hopper-evolution}
\end{figure}

Each non-Byzantine node processes roughly 200,000 identifiers before \(T_a\). In Figure~\ref{fig:hopper-evolution}, saturated Hopper remains highly polluted and cumulative AUPE adapts slowly. At the same 500-byte budget, Hopper-D rapidly reduces the initial spike because decay makes the post-attack frequency ratio influential.

\begin{figure}[!t]
  \centering
  \includegraphics[width=0.7\linewidth]{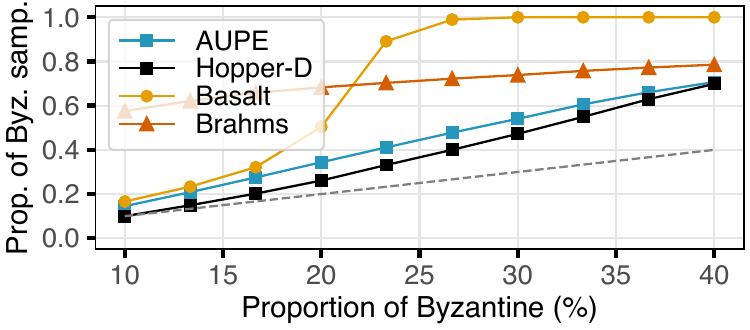}
  \caption{Average pollution over rounds 11,000-20,000}
  \label{fig:hopper-convergence}
\end{figure}

Hopper-D remains below BRAHMS and BASALT and at or below AUPE across the evaluated Byzantine fractions (Figure~\ref{fig:hopper-convergence}). AUPE's average approaches Hopper-D at high \(f\) because AUPE decreases gradually over the measurement window, not because it has already stabilized. Figure~\ref{fig:hopper-evolution} shows that Hopper-D reaches its steady regime within a few hundred rounds, while AUPE remains more polluted through most of the post-attack interval.
Hopper-D reaches the target pollution at \(f=10\%\), but pollution is approximately 0.26, 0.47, and 0.70 for \(f=20\%,30\%,40\%\). Thus, freshness improves resilience but neither guarantees uniform sampling at high \(f\) nor makes 500 bytes universally sufficient.

\subsection{RQ3: What Does Trusted Collaboration Add?}

\begin{figure}[!t]
  \centering
  \includegraphics[width=\linewidth]{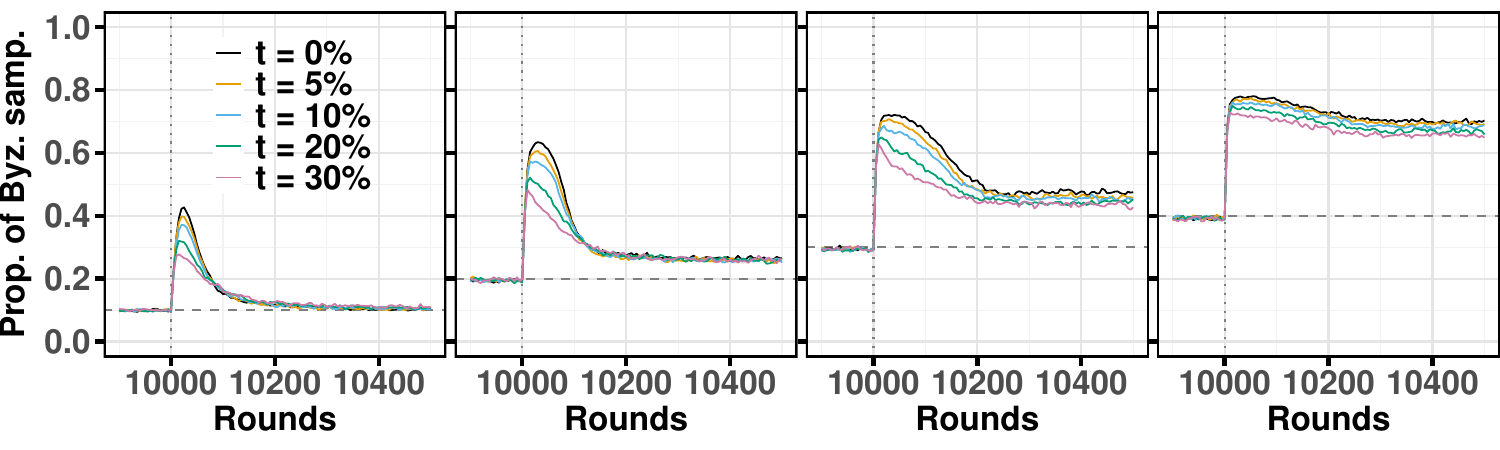}
  \caption{Hopper-D pollution around \(T_a\) as trusted fraction \(t\) varies. Columns show \(f=10\%,20\%,30\%,40\%\).}
  \label{fig:merge-onset}
\end{figure}

We run Hopper-D with \(t\in\{5\%,10\%,20\%,30\%\}\). Each trusted node keeps ten authenticated trusted peers, selects one uniformly each round, executes the pairwise averaging and deterministic reconstruction from Algorithm~\ref{alg:trusted-merge}, and applies the trusted debiasing rule.
Increasing \(t\) consistently lowers the transient pollution peak, while steady curves remain closer together (Figure~\ref{fig:merge-onset}).

Relative to \(t=0\), \emph{steady gain} is the pollution reduction averaged over non-Byzantine nodes and rounds 11,000--20,000. \emph{Peak gain} is the reduction with not trusted nodesright after the attack when the pollution is at its peak, all configurations use the same round.

Peak gains range from 1.5\% to 38\%, increase with \(t\), and generally decrease with \(f\) (Figure~\ref{fig:merge-gains}(a--b)). Steady gains range from \(-5.3\%\) to 9.7\%, negative values occur at low \(f\) with large \(t\) and represent under two percentage points of absolute pollution. Collaboration therefore primarily mitigates the transient and is not uniformly beneficial in steady state.

\subsection{RQ4: Does Collaboration Reveal Trusted Nodes?}

\begin{figure}[!t]
  \centering
  \begin{subfigure}{0.325\linewidth}
    \includegraphics[width=\linewidth]{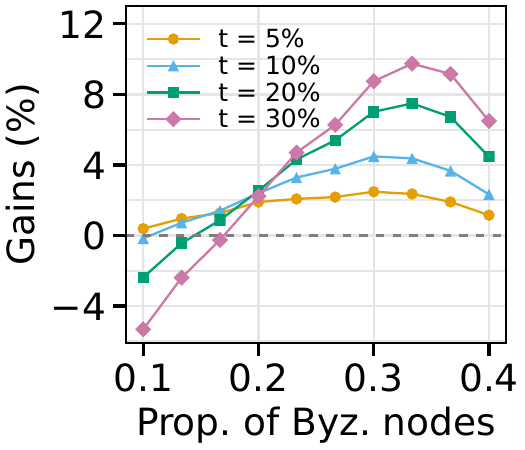}
    \caption{Steady gain}
  \end{subfigure}
  \begin{subfigure}{0.325\linewidth}
    \includegraphics[width=\linewidth]{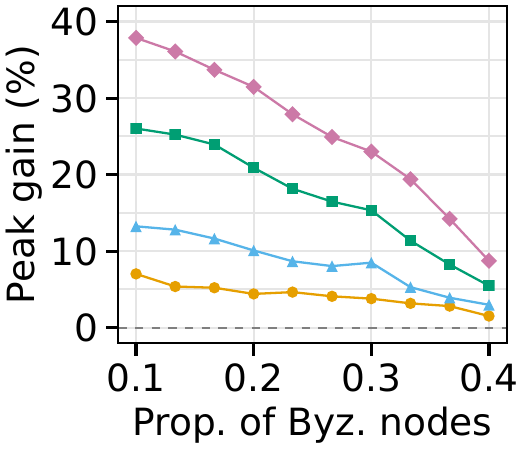}
    \caption{Peak gain}
  \end{subfigure}
  \begin{subfigure}{0.325\linewidth}
    \includegraphics[width=\linewidth]{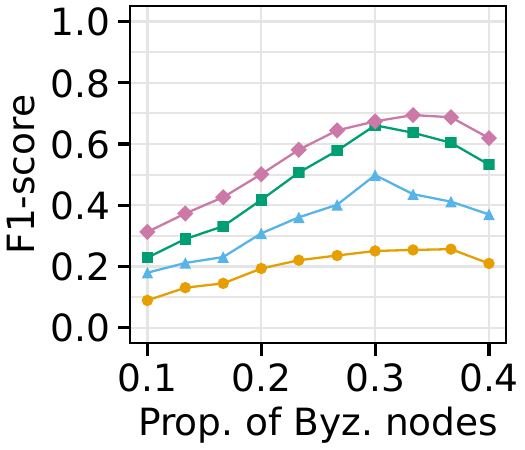}
    \caption{Trusted-node \(F_1\)}
    \label{fig:trustedidentification}
  \end{subfigure}
  \caption{Trusted Hopper-D: steady gain, peak gain, and trusted-node re-identification relative to Hopper with no trusted nodes.}
  \label{fig:merge-gains}
\end{figure}

We give the adversary every non-Byzantine view during rounds 10,000--10,049 and knowledge of its Byzantine identifiers. It clusters each view's Byzantine fraction with \(K=2\), labels the lower-centroid cluster as trusted, and compares its predictions with ground truth. Figure~\ref{fig:trustedidentification} reports \(F_1\).
For \(t=5\%\), precision remains below 0.15 despite recall near 0.95, yielding \(F_1\le0.26\).
For \(t=10\%\), precision remains below 0.35 and \(F_1\le0.50\). Sparse deployments therefore produce too many false positives for precise targeting.
Re-identification becomes partial at larger \(t\).
For \(t=20\%\), precision reaches 0.55, recall 0.82, and \(F_1=0.66\), while for \(t=30\%\), precision reaches 0.65, recall 0.75, and \(F_1=0.70\).
Even then, 35-45\% of predicted targets are false positives.
The large \(t=30\%\) case exposes the approach's limit rather than an expected deployment.





\section{Related Work}
\label{sec:related-work}

\textbf{Peer sampling and Byzantine resilience.}
Cyclon and Newscast~\cite{voulgarisCYCLON2005,tolgyesi2009adaptive} build random-looking overlays but assume compliant peers, PeerSwap~\cite{peerswap:2024} proves convergence-time bounds for random neighborhoods by swapping peer positions over a fixed graph. Hopper instead targets adversarially biased identifier streams and neither preserves a fixed topology nor proves uniform convergence.

Byzantine-resilient designs constrain influence differently. Secure Peer Sampling~\cite{sps:2010} and SecureCyclon~\cite{Antonov:2023} detect and exclude deviating nodes, BRAHMS~\cite{Bortnikov:2008} bounds push/pull influence and protects a history subview, whereas BASALT~\cite{basalt:2023} applies seeded min-wise sampling to the complete view. Honeybee~\cite{honeybee:2025} combines shared secure randomness, verifiable random walks, and table consistency checks to obtain near uniform samples and expose equivocation.
Hopper verifies neither paths nor identities. Instead, Hopper downweights estimated overrepresentation in BRAHMS's fresh streams. It assumes a Sybil-resistant membership layer and does not claim Honeybee's verifiable or Sybil-resilient guarantees.

Recent application specific designs optimize other neighbor selection objectives.
DISC-NG~\cite{discng:2024} protects bounded Ethereum service-advertisement caches using DHT routing and signed waiting-time tickets, Constellation~\cite{constellation:2026} derives a low-degree, diameter-two overlay from an FBA quorum system, and LIFT~\cite{lift:2025} uses cryptographically secure pseudorandomness to protect Elevator's hub selection~\cite{elevator:2024}.
BLADE~\cite{blade:2025} instead secures decentralized learning by robustly aggregating model updates under heterogeneous data and Byzantine behavior. These systems target service discovery, FBA communication, hub formation, or learning-layer aggregation, not debiasing per-round push/pull identifier streams.


RAPTEE and AUPE are Hopper's direct parents. RAPTEE uses TEE-backed nodes to help repair polluted views~\cite{Pigaglio:2022}, AUPE adds occurrence-aware cleaning and averaging of exact tracking state~\cite{aupe:2024}. Hopper retains AUPE's frequency signal but adds fixed-budget fingerprint estimation, decay under capacity pressure, and fingerprint-aware trusted reconstruction.

\textbf{Uniform sampling from biased streams.}
Anceaume et al. construct exact uniform node sampling from adversarially biased streams~\cite{anceaume2013uniform} relying on Count Min Sketches that we have shown are not discriminative enough for filtering node identifiers in long-lived peer sampling protocols.

\textbf{Frequency estimation under skew.}
Count-Min Sketch provides fixed-memory, non-negative estimates and a linear, mergeable representation~\cite{cormode:2005}.
Conservative Update, Count-Mean-Min, and Lossy Conservative Update mitigate collision error through different policies~\cite{estan:2003,deng:2007new,goyal:2011,goyal:2012}.
Cold Filter separates cold and hot items, XY improves cold-item estimates through identifier decomposition, and BitMatcher adapts widths while retaining fingerprints~\cite{coldfilter:2019,liu:2023,bitmatcher:2024}.
Hopper evaluates these designs by distribution fidelity, class separation, and adversarial bias rather than generic point-query error.

\textbf{Aging unbounded streams.}
Exponential histograms provide exact-window statistics, forward decay weights observations by age, and TinyLFU periodically ages insertion frequencies~\cite{exphist:2002,forwarddecay:2009,tinylfu:2017}.
Hopper-D instead uses saturation-triggered halving and provides neither window queries nor time-based decay, but makes recent ratios dominate a benign prefix.

\textbf{Merging distributed summaries.}
Mergeable summary theory formalizes composition without losing approximation guarantees~\cite{cmsmerge}.
BitMatcher lacks these semantics because bucket states change entry locations and reconstruction may evict entries, Hopper therefore uses authenticated, fixed-budget lossy reconstruction rather than an unbiased union or sum.


\section{Conclusion}
\label{sec:conclusion}

In this paper, we studied how cumulative frequency debiasing is insufficient for long-lived Byzantine peer sampling, and propose Hopper that combines BitMatcher with BMDecay, which preserves recent frequency contrast. 
Our approach reduces, with fixed memory, delayed-attack pollution relative to existing solutions.
Trusted collaboration reduces transient peaks, but its steady benefit is parameter-dependent.
Hopper establishes freshness as distinct from bounded memory and guarantees neither uniform sampling at high Byzantine fractions nor complete trusted-node anonymity.

The broader lesson is that adversarial peer sampling is an online adaptation problem, not only a counting problem as investigated in its foundational works.
For frequency-based debiasing, an estimator is useful if it exposes
temporally relevant information and allows the protocol to discriminate
between behaviors when they exist.
Conversely, an approximate summary that ages observations can outperform exact cumulative state after an adversarial distribution shift.
This result motivates designing and evaluating peer-sampling protocols under
non-stationary attacks, using metrics such as information freshness, adaptation delay, pollution, and recovery.

Our work opens three research directions.
First, adaptive aging should be further studied with regards to bounds that relate memory, adaptation delay, and view pollution under churn and adaptive attacks.
Second, collaboration should propagate evidence without making trusted nodes statistically distinguishable to the adversary, including active inference rather than only passive ones.
Third, Hopper's approach should be composed with applications that rely on peer sampling to actually measure whether lower view pollution improves
application-level key performance and actual robustness.



\bibliographystyle{IEEEtran}
\bibliography{references}

\end{document}